%% file: Main_v0.4.tex
\documentclass[%
aip,
 amsmath,amssymb,
reprint,%
]{revtex4-1}

\usepackage{graphicx}
\usepackage{dcolumn}
\usepackage{lipsum}
\usepackage{bm}
\usepackage{tabu}
\usepackage[utf8]{inputenc}
\usepackage[T1]{fontenc}
\usepackage{mathptmx}
\usepackage{color}
\usepackage[export]{adjustbox}
\usepackage{siunitx}
\usepackage{hyperref}

\newcommand{\red}[1]{\textcolor{black}{ #1}}

\begin{document}
\preprint{AIP/123-QED}

\title{Structured Light's Applications: A perspective}
\author{Carmelo Rosales-Guzm\'an}\email{carmelorosalesg@cio.mx}
\affiliation{Centro de Investigaciones en Óptica, A.C., Loma del Bosque 115, Colonia Lomas del campestre, C.P. 37150 León, Guanajuato, Mexico.}%

\author{Valeria Rodr\'iguez-Fajardo}
\affiliation{Departamento de F\'isica, Universidad Nacional de Colombia Sede Bogotá,  Carrera 30 No. 45-03, Bogot\'a, Colombia} 

\date{\today} 
\begin{abstract} 
For the past few decades, structured light has been gaining popularity across various research fields. Its fascinating properties have been exploited for both previously unforeseen and established applications from new perspectives. Crucial to this is the several techniques that have been proposed for both their generation and characterisation. On one hand, the former has been boosted by the invention of computer-controlled devices, which combined with a few optical components allow flexible and complete control of the spatial and polarisation degrees of freedom on light, thus enabling a plethora of proof-of-principle experiments for novel and old applications. On the other hand, characterising light beams is important not only for gaining better insights into light's properties but also for potentially being used as metrics. In this perspective, we thus offer our take on a few key applied research fields where structured light is particularly promising, as well as some pivotal generation and characterisation techniques. In addition, we share our vision of where we believe structured light's applications are moving towards.
\end{abstract}

\maketitle

\section{Introduction}
Structured or complex light is emerging as one of the most promising areas of modern optics at both the fundamental and applied levels \cite{Forbes2021StructuredLight,Roadmap}. In the early years, research mostly focused on generating scalar beams, characterised by a homogeneous transverse polarisation distribution, where only the spatial intensity and phase distributions were manipulated. Then, the field started evolving towards more complex light states, where other degrees of freedom are also tailored \cite{Shen2022}. Of particular interest are complex vector light beams, featuring a non-separable superposition of the spatial and polarisation degrees of freedom thus giving rise to light beams with non-homogeneous transverse polarisation distributions \cite{Dieter1972,Mushiake1972,Shen2022}. These are sometimes called ``classically-entangled'' modes due to their mathematical similarity to entangled photons \cite{forbes2019classically,toninelli2019concepts}, however, this term is highly controversial and thus should be used with precaution\cite{Karimi2015}. Nowadays there are all sorts of techniques for the generation\red{\cite{Tidwell1990, Niziev2006, Passilly2005, Marrucci2006, Berkhout2010, Mirhosseini2013, Naidoo2016, Devlin2017, Radwell2016, Kozawa2005}} and characterisation of both scalar\red{\cite{Bryngdahl1973, Soskin1997, Vasnetsov2003,  Leach2002, Leach2004, ShikangLi2018, kulkarni2017, Mazilu2012, Otte2022}} and vector\red{\cite{McLaren2015, Ndagano2016, Liu2017, Yang2018, Otte2018, Bhebhe2018a, Selyem2019, Zhu2019, Aiello2022}} beams. In particular, the latter has been realised experimentally in cylindrical, elliptical, and parabolic coordinates. Examples of these are Laguerre- and Bessel-Gauss \cite{Zhan2009,Dudley2013,Otte2018b,Ren2015,Medina-Segura:2023}, Mathieu- and Ince-Gauss \cite{Rosales2020,Yao-Li2020,Otte2018a}, parabolic and accelerating  \cite{ZhaoBo2022,HuXiaobo2021}, as well as abruptly auto-focusing vector beams\cite{Hu2022abruptly,Hu2023}. 

With more mature generation and characterisation techniques, and given the structured beams' fascinating properties, researchers are engaging in the development of pioneering applications in fields such as optical manipulation\red{\cite{Bhebhe2018,yang2021}}, high-resolution microscopy\red{\cite{Lukinavicius2024}}, optical metrology\red{\cite{Toppel2014,Hu2019,Rosales2013SR,Belmonte2015}}, and classical and quantum communications\red{\cite{Ndagano2018,Li2016,Ndagano2017,Milione2015e,Wang2015}}, amongst many others\cite{Rosales2018Review}. Nevertheless, there is still room for developing generation techniques tailored to the demands of applications and characterisation methods that can be used as metrics in potential applications. As such, in what follows, we will provide a short description of current generation and characterisation techniques, followed by an overview of some of the applications developed using structured light. We will finalise by giving our perspective on the future development of structured light generation and characterisation techniques, as well as possible applications, some of which are probably on their way, as we write this perspective. 

\section{Generation and Characterisation}
This section provides a broad overview of generation and characterisation techniques. Initially, structured beams were produced using optical elements such as axicons, \red{which can be traced back to the 1950s}\cite{McLeod1954,Indebetouw1989},  photographic-printed holograms, \red{which became popular in the 1990s}\cite{Bazhenov1992}, spiral phase plates\cite{Oemrawsingh04} \red{dating back to the 2000s}, and more recently with birefringent waveplates, such as q-plates\cite{Marrucci2006}, \red{developed in 2006,} and metasurfaces\cite{Devlin2017} \red{from the 2010s}. In addition, the use of more elaborated flat optical elements, utilising metasurfaces \cite{Dorrah2022} or liquid crystals \cite{ChunYu2023} are also gaining popularity, as they have the potential to replace bulky elements with compact ones, as desired for real-world applications. The direct generation of structured light beams from a resonator cavity has also been explored and could represent an alternative for various applications\cite{Ngcobo2013,Naidoo2016,Forbes2019Lasers}. 
\red{Initially developed in the 1970s within the projection technology,} two key devices that have been crucial in the demonstration of novel structured light beams and applications' proof-of-concept experiments are LCoS spatial light modulators (SLMs) \cite{SPIEbook} and digital micromirror devices (DMDs) \cite{Gong2014, Mitchell2016, Scholes2019, hu2021generation,  Hu2022}, as they offer unprecedented versatility and flexibility. As such, this perspective focuses on these since they will continue leading the structure light field in the above-mentioned aspects. 
\begin{figure*}
    \centering
    \includegraphics[width=.98\textwidth]{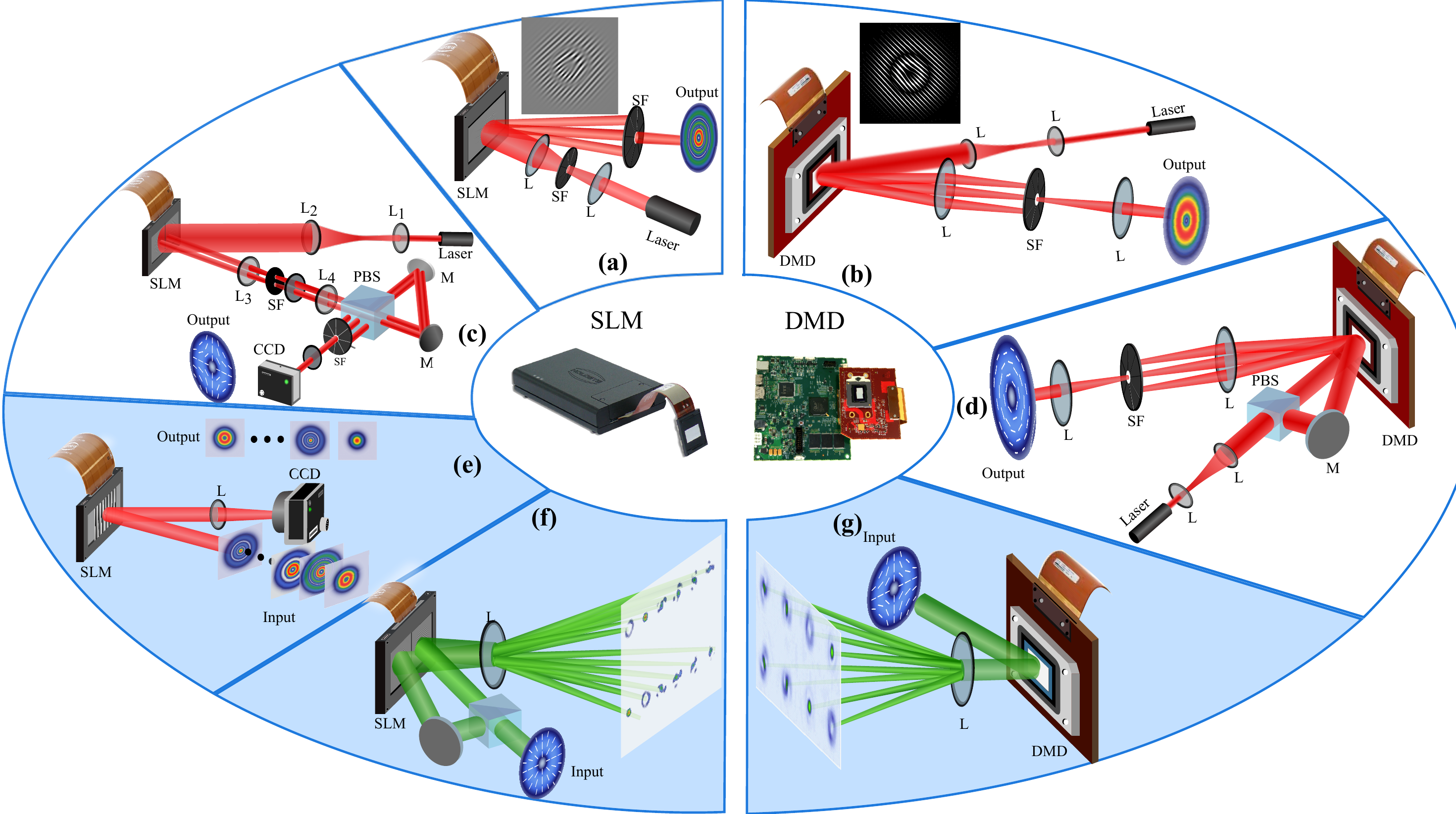}
    \caption{Schematic representations of experimental setups for the generating scalar beams using (a) an SLM and (b) a DMD. The insets show the computer-generated holograms for each case, a grey-scale for an SLM, and a binary for a DMD. Typical experimental setups for generating vector beams using (c) an SLM and (d) a DMD.  (e) Characterisation of scalar beams through modal decomposition. Characterization of vector beams through the vector quality factor implemented with (f) an SLM and (g) a DMD. }
    \label{fig:generationcharacterization}
\end{figure*}

To begin with, it is worth reminding that neither \red{SLMs} nor DMDs allow natively manipulating the amplitude and phase simultaneously. As such, various techniques have been developed to manipulate both indirectly. In the case of LCoS SLMs, only the phase can be modulated directly by displaying a grey-scale image on the SLM's screen \cite{SPIEbook}. Hence, several techniques have been developed to modulate the amplitude as well, being complex amplitude modulation (CAM), \red{demonstrated for the first time in 2007,} one of the most popular, as it produces high-quality light beams\cite{Arrizon2007}. A typical experimental setup to generate arbitrary scalar beams is illustrated in Fig.~\ref{fig:generationcharacterization} (a), where an example image of a computer-generated hologram is also shown (see \cite{SPIEbook} for further details). On the other hand, DMDs can only modulate the amplitude using binary images, therefore the amplitude and phase of a complex light field need to be encoded onto binary holograms. To this end, the phase and amplitude information is encoded in the first diffraction order of a periodic binary amplitude grating\cite{lee1979,Lee:74}, whose theory was developed in the mid-1960s \cite{brown1966}.  
A typical experimental setup to generate scalar beams is schematically illustrated in Fig.~\ref{fig:generationcharacterization} (b), where an example binary computer-generated hologram is also shown (see for example \cite{Hu2022} for further details).

In contrast to scalar beams, the techniques to generate vector beams require the simultaneous manipulation of the spatial and polarisation degrees of freedom. \red{Since their beginnings in the 1970s,} various methods have been proposed, which can be classified into two main categories: geometric- and dynamic-based phase modulation. The former is based on the direct conversion of spin to orbital angular momentum, achieved either through liquid crystal wave-plates, such as q-plates\cite{Marrucci2006}, or metamaterials\cite{Devlin2017}. The latter can be achieved by interferometric means, or via a sequential manipulation of the spatial profile of both polarisation components. Notably, the use of SLMs\cite{Davis2000,Moreno2012,Mitchell2017,Rong2014} and DMDs  \cite{Mitchell2016,Scholes2019,Gong2014,hu2021generation,Hu2022} allows the generation of scalar beams with arbitrary spatial shapes, that in combination with polarisation optics enable the generation of a wide variety of vector beams. Fig.~\ref{fig:generationcharacterization} (c) illustrates schematically a Sagnac-based interferometric generation of vector beams using an SLM (see \cite{Perez-Garcia2017} for a detailed description). Similarly, Fig.~\ref{fig:generationcharacterization} (d) illustrates a polarisation-insensitive experimental setup for the generation of arbitrary vector beams (see\cite{Rosales2020} for details). Other approaches focused on generating vector beams that are highly stable\cite{RodriguezFajardo2024}, which is necessary for applications that require monitoring over time.
 
A proper characterisation of structured light fields is also of great relevance since it provides insightful information about many of their properties, which, in turn, can potentially be used as metrics in applications. While various techniques have been proposed for scalar beams, there are not as many for vector beams. In the first case, many techniques focus on determining their Orbital Angular Momentum (OAM) spectrum. Some rely on interferometric methods, where the beam of interest is made to interfere with a plane wave or an inverted copy of itself\cite{Soskin1997,Bryngdahl1973}. Others rely on the far-field diffraction pattern produced by the beam after traversing an aperture with a given symmetry \cite{deAraujo2011,Mazilu2012}. The use of refractive optical elements to transform the azimuthal phase of OAM beams into transverse phase gradients, widely known as mode sorters, has also become popular \cite{Lavery2012,Fontaine2019}. Amongst these, modal decomposition is perhaps one of the most powerful, as it provides the complete reconstruction of an unknown field. The key idea behind modal decomposition is that any field can be expanded in terms of the basis elements of an orthonormal complete set of beams.
Further, the expansion coefficients can be unambiguously determined through the inner product of the unknown field and the basis elements, which can be implemented by displaying the complex conjugate of the basis function as a digital hologram onto an SLM\cite{Flamm2012,Schulze2012,Pinnell2020}. The modal amplitude is then measured in the far-field, as the on-axis intensity of the output beam. Importantly, this procedure also allows measuring the modal phase thus completely determining the unknown complex light field. This is schematically represented in Fig.~\ref{fig:generationcharacterization} (e), where different input beams projected onto the same hologram, produce different far-field distributions. 

Regarding the characterisation of vector beams, only recently are researchers becoming aware of the necessity of developing metrics for their properties, thus a few have been proposed so far. Along this line, Stokes polarimetry provides a direct way to reconstruct their transverse polarisation distribution through a minimum of four intensity measurements\cite{Goldstein2011}. Such intensities are typically measured sequentially, one by one, although it is also possible to measure all simultaneously by taking advantage of the polarisation-independent nature of DMDs \cite{Zhao2019}. This enables the real-time reconstruction of polarisation, paving the way for applications in optical metrology where the system under inspection evolves dynamically over time. To achieve this, the DMD displays a multiplexed digital hologram, splitting the input vector beam into four identical copies propagating along different paths. Each beam is then passed through the required polarisation filters and all are recorded simultaneously by a CCD, from which the polarisation is reconstructed after some simple manipulation. Even though Stokes polarimetry is a powerful tool that allows a two-dimensional mapping of the transverse polarisation distribution, and in principle contains all necessary information, its use as a metric is not straightforward. Therefore additional characterisation parameters that could be used as indicators are needed.

At this stage, the mathematical similarity between vector modes and quantum entanglement becomes relevant. More precisely, in the context of quantum mechanics, the degree of entanglement between two photons can be measured through a quantity known as {\it Concurrence (C)}. Such quantity \red{was} adapted \red{2015} to vector modes \red{as a measure of} their purity, or in other words, the degree of coupling or "classical entanglement" between their spatial and polarisation degrees of freedom. To differentiate it from the quantum cases, it has received the name Vector Quality Factor (VQF) and it assigns values of 1 and 0 to pure vector and scalar modes, featuring maximum and null degrees of non-separability, respectively\cite{McLaren2015}. 
Experimentally, the VQF can be measured using the computer-controlled devices mentioned above. The use of SLMs requires splitting the vector beams into their right- and left-handed polarisation components, which are projected afterward onto a series of six spatial filters encoded as digital holograms on the device \cite{Ndagano2016,Bhebhe2018a}. Each combination of polarisation and spatial projection provides a unique far-field intensity pattern, twelve in total, at the focal plane of a lens, where only the normalised on-axis intensity value of each combination is measured. Crucially, all can be determined simultaneously through a multiplexing approach, as schematically illustrated in Fig.~\ref{fig:generationcharacterization} (f). The use of a polarisation-insensitive DMD enables to determine the VQF directly without the need to split it into its polarisation components thus reducing the number of required measurements to a minimum of 8 \cite{Zhaobo2020,Manthalkar2020}. In this approach, the vector beam is first projected onto the orthogonal spatial modes that constitute the vector modes and then traced over the polarisation degree of freedom implemented with polarisation filters. Fig.~\ref{fig:generationcharacterization} (g) illustrates schematically a possible experimental implementation of this approach. Crucially, the VQF can also be computed directly by integrating the Stokes parameters over the whole transverse plane in a basis-independent manner, avoiding the need for \textit{a priori} knowledge of the spatial modes constituting the vector mode, a technique that represents a notable advance in characterising complex vector light fields \cite{Selyem2019}. This technique was recently used to demonstrate the invariance of vector beams upon propagation through complex media\cite{nape2022}.

\section{Applications}
Structured light beams are paving the path towards novel applications in various research fields, some with a history dating back to the early years of structured light and others still at an early development stage. Importantly, applications inspired by the similarities between classical and quantum entanglement have also been proposed, representing a research line on its own. Here, we will briefly comment on applications in some fields with potentially higher impact, namely optical communications, optical manipulation, and optical metrology. 
 \begin{figure*}
    \centering
    \includegraphics[width=.99\textwidth]{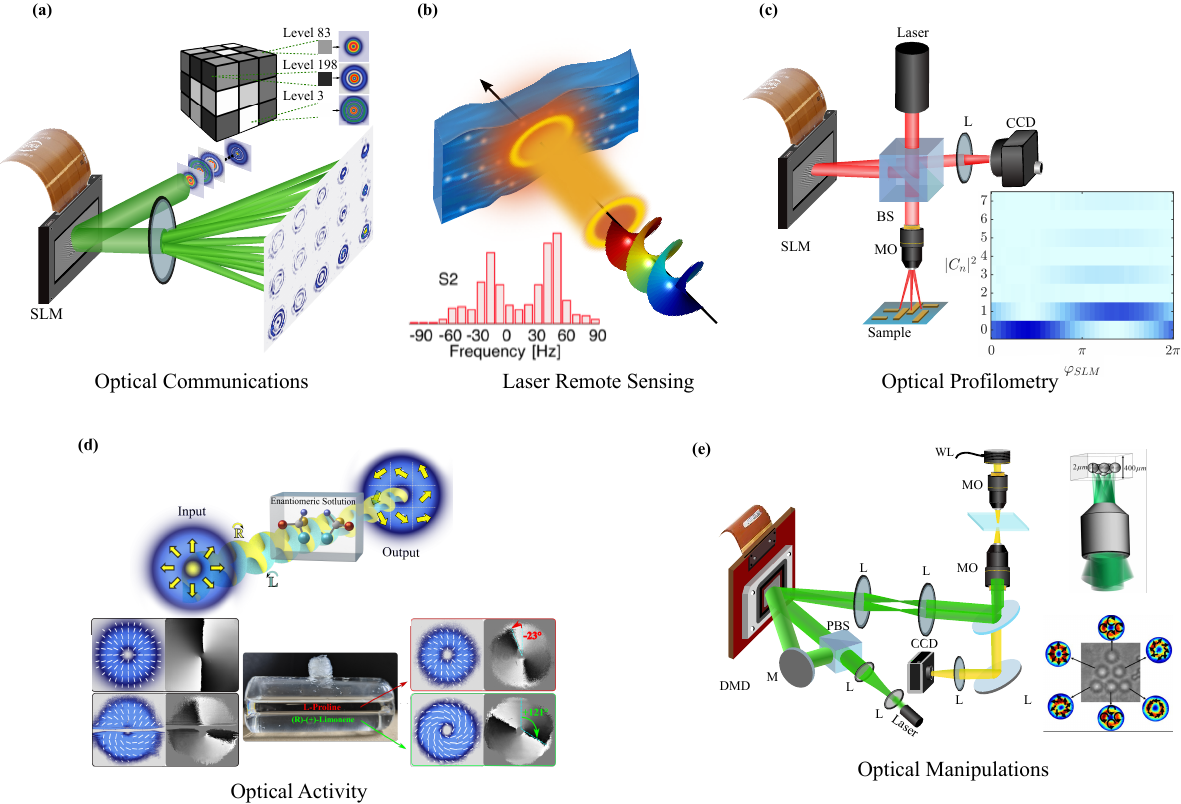}
    \caption{Applications of structured light beams in (a) Optical communications; (b) laser remote sensing; (c) optical profilometry\cite{RodriguezFajardo2022}, Adapted with permission from
    \href{https://doi.org/10.1103/PhysRevApplied.18.064068}{Rodríguez-Fajardo \textit{et al.}, Phys. Rev. Appl. 18, 064068 (2022)}, Copyright 2022 American Physical Society; (d) optical activity\cite{Hu2024}, Adapted with permission from Hu \textit{et al.}, ACS Photonics (2024) Copyright 2024 American Chemical Society; (e) Optical tweezers\cite{Bhebhe2018}.}
    \label{fig:applications}
\end{figure*}

\subsection{Applications in Optical Communications}
It has been more than sixty years since the seed for a global-scale communication network was planted \cite{kao1966}. Nowadays, it is almost impossible to conceive a world without the Internet, which has become essential in our everyday activities, such as remote work, bank transactions, social media, entertainment, and online shopping, amongst others. Key to this was the demonstration of the first working laser and the invention of the silica-based optical fibres, which led to the realisation \red{in 1965} that information could be encoded in the form of a laser beam and transmitted through optical fibres. In turn, this ignited a revolution in telecommunications that laid the foundations of our current communications systems, changing forever the way we share information. Over the years, the global capacity demand has been increasing at an exponential rate, inspiring some of the greatest minds in the field to come up with practical solutions or \red{ingenious} coding schemes for increasing the capacity of data transmission. This, for example, led to breakthrough technological advancements, such as the development of low-loss single-mode fibres, erbium-doped fibre amplifiers, high-spectral-efficiency coding, amongst others \cite{Richardson2013}. In addition, various properties of light have been exploited for information transmission in multiple independent channels. For example, in wavelength-division multiplexing, multiple channels are defined by unique wavelengths. Light, however, \red{possesses} other degrees of freedom, such as polarisation, amplitude, phase, and time, all of which have also been exploited. Despite all the aforementioned technological advances, the capacity of our current communications systems is reaching a limit, associated mainly with nonlinear effects in optical fibres, leading to a “capacity crunch'', which is within sight \cite{Richardson2010}. Therefore, the need for innovative communication systems with higher transmission capacity and desirably higher security has become evident. 

Amongst the potential solutions that have been proposed, space division multiplexing (SDM) stands out as one of the most promising alternatives \cite{Richardson2013,Li2014}. Of particular interest is mode division multiplexing (MDM), suggested almost as early as the invention of the optical fibre, a technology with the potential to address future bandwidth issues \cite{Berdague1982}. In MDM-based optical communication systems, each spatial mode, from an orthogonal modal basis, is used as an independent data carrier, thus increasing the overall capacity by a factor equal to the number of modes used \red{as demonstrated in 2004} \cite{Gibson2004}. While theoretically, the number of usable modes is infinite, in practice this number is limited by technical constrictions, such as beam divergence or mode degradation \cite{Zhao2015}. Different bases for MDM both in free-space and optical fibre communications have been proposed. Along this line, spatial modes with orbital angular momentum (OAM), usually Laguerre-Gaussian modes, have received special attention as they can be easily generated and detected with phase-only optical elements\cite{Gibson2004,Willner2015,Bozinovic2013}. Crucially, the use of both transverse parameters of Laguerre-Gaussian modes, azimuthal (related to the OAM) and radial, allows forming a densely packed mode space, therefore further increasing the data transmission rate. In this way, the number of information channels increases considerably in proportion to the radial index $p$, since each OAM mode defines an infinite number of modes with different radial numbers \cite{Trichili2016, Xie2016}. A schematic representation of this application is illustrated in Fig.~\ref{fig:applications} (a). In addition, vector beams, non-separable in their spatial and polarisation degrees of freedom, have also recently gained popularity \cite{Ndagano2018} since they are more resilient to perturbations in one-sided channels (where only one of the two degrees of freedom is perturbed), and it is possible to implement communication systems with higher security \cite{Ndagano2017,Otte2020}, as well as novel information encoding schemes \cite{Singh2023}. Importantly, current encoding techniques can also be combined with MDM, to increase the bandwidth of available communication systems by a factor proportional to the number of spatial modes employed.

\subsection{Applications in Optical Metrology}
Optical metrology is an expansive area of science and technology encompassing all measuring techniques that use light as the main probing tool. Optical approaches are usually preferred over traditional mechanical ones since they provide non-contact measurements with high precision, speed, and flexibility. They have been used to measure the physical properties of solids and fluids, such as their geometrical properties, chemical composition, temperature, pressure, and even distances and velocities \cite{OpticalMetrology}. By way of example, in this section, we will briefly comment on applications from three fields: laser remote sensing, surface metrology, and molecular chirality sensing. These constitute only a small sample of the many more that have been proposed or are currently under development but we hope they will give a sense of the great potential structured light provides.

Laser remote sensing encompasses a variety of techniques commonly used to monitor the location and velocity of moving targets, with applications in fields as diverse as medicine, astronomy, meteorology, and aeronautics. For many of them, the longitudinal Doppler shift of light, i.e. the perceived change in frequency of waves caused by the relative motion between a transmitting source and a detector, is crucial \cite{Measures1992}. In essence, the light reflected or scattered back by a moving target experiences a frequency shift proportional to the target's speed\red{, such that} the frequency shift measurement enables determining the velocity of the target. The Doppler effect, however, allows measuring only the longitudinal velocity, making the transverse velocity component undetectable. Nonetheless, having access to the full velocity vector is crucial for many applications. Therefore, several techniques rely on indirect ways, such as using multiple light beams or the fast mechanical realignment of the direction of propagation of a single laser beam, making its implementation rather complicated. Other techniques, such as laser Doppler anemometry, mainly used to study fluid dynamics, rely on interference patterns produced by the interference of two coherent beams.

In 2011, Belmonte and Torres proposed a theoretical approach that provided the means to directly measure the transverse velocity component, using light beams with non-homogeneous transverse phase profiles \cite{Belmonte2011}. The technique relies on the fact that transverse phase gradients in optical beams introduce additional energy flows in the direction perpendicular to propagation. Hence, the phase of the light reflected or scattered back from a target moving along the plane perpendicular to the illuminating beam, of a size compared to the trajectory, contains information about its position and velocity. The velocity information is then extracted from the Fourier transform of the intensity signal produced from the coherent interference of the backscattered light and a reference beam. If the trajectory of the target is known {\it a priory}, the structure of the phase can be adapted accordingly to simplify the velocity measurement \cite{Rosales2013SR}. For example, obtaining the rotational velocity of targets moving in circles is easier if illuminated with an azimutally-varying phase. 

Crucially, it is possible to elucidate advanced systems with the capability of measuring simultaneously all velocity components in a 3D motion, longitudinal and transverse. In a first approximation, a target describing a 3-dimensional helical trajectory is illuminated sequentially with a Gaussian and a vortex beam, such that the former and the latter are used to determine the longitudinal and rotational velocity components, respectively\red{, as demonstrated in 2014} \cite{Rosales2014OE}. In a more complex approximation, the targets were illuminated with vector beams constituted by the non-separable superposition of a Gaussian beam and a vortex beam with orthogonal polarisation states\cite{Hu2019}. Here, the orthogonality in polarisation allows for simultaneously determining both velocity components. The transverse Doppler shift has also been used to measure directly the vorticity of fluids\cite{Belmonte2015}. In this technique, a fluid flux is illuminated with a vortex beam, as schematically depicted in \red{ Fig.~\ref{fig:applications} (b)}, and the returned signal produces a frequency spectrum (see bottom inset) whose centroid is directly linked to the vorticity. 

Surface metrology is integral to many industrial and scientific applications \cite{Leach2015}. For instance, for nano-manufacturing evaluation and vertical standards calibration, the measurement of height jumps is fundamental \cite{Brand1995,Dai2005,Fang2017}. In general, due to their non-contact nature and enhanced speed \cite{Leach2011,Tan2013,Wu2017a,Marrugo2020}, optical approaches \cite{Creath1987,Tiziani1994,Knauer2004,Huang2018,Kim2010,RodriguezFajardo2021} are preferred to their tactile counterparts \cite{Sasagawa2017,Bennett1981,Binnig1982,Binnig1986,Koenders2006}. Their performance, however, can be affected by system instabilities, arising from, for instance, vibrations \cite{deGroot2015}, and the need for a reference optical flat for reflective samples. As such, researchers have turned their attention to structured light in \red{the} quest for more robust techniques. In particular, the far-field scattered light from a cliff-like feature when illuminated with a Hermite-Gaussian (HG) mode has been theoretically explored to determine its height and side-wall angle\cite{Dou2022}. Another popular approach is mode projection, where the height of step-like features is determined from the modal content (in a specific beam family basis) of the reflected light from the step. There have been investigations based on orbital angular momentum modes \cite{Torner2005,MolinaTerriza2007}, modified Gaussian modes\cite{Hermosa2014}, and HG modes\cite{RodriguezFajardo2022}\red{, approaches that were demonstrated in 2007, 2014 and 2022, respectively}. Fig.~\ref{fig:applications} (c) presents a schematic drawing of the optical setup of the latter and the inset shows experimental results of the modal decomposition of a digital step of \SI{60}{\nano\meter} as a function of a constant step-like phase term added to the HG-projection hologram. In this particular case, the height is determined only using the zeroth and first modal coefficients and the accuracy mainly depends on the number of phase terms measured.

Structured light has a promising future in optical activity sensing, a phenomenon commonly associated with the interaction of polarised light and enantiomers, such as optical rotation and circular dichroism. Enantiomers are pairs of chiral molecules, one being the mirror image of the other, identical in most regards but distinguishable in their interaction with other chiral objects. Crucially, the human body makes amino acids and sugars, the molecular building blocks of life, using only the right-handed variety. This intriguing aspect has great relevance in the food and drug industry, since, for instance, two enantiomers can have very different physiological effects\red{\cite{sanganyado2017}}. As such, there is great interest in developing techniques capable of discriminating between enantiomers. Traditionally, most existing approaches rely on polarised light even though in the last decade it has been demonstrated that the orbital angular momentum of light can also engage with chiral molecules \cite{Rosales-Guzman2012, KaynForbes2019, Brullot2016}. This is a topic that is gaining increasing popularity, see for example \cite{KaynForbes2021} for an interesting review. Along this line, a vectorial version proposed \red{in 2024}, takes advantage of the non-homogeneous polarisation distribution of vector beams \cite{Hu2024}. In this novel approach, the position-dependent polarisation direction of the vector beam across the transverse plane enables the measurement of the concentration variations across different sample regions. The technique relies on a spatially resolved measurement of the intermodal phase of the probe vector beam, which depends on parameters specific to the sample and its concentration. As schematically illustrated in the top inset of \red{ Fig.~\ref{fig:applications} (d)}, an input vector beam with a well-known polarisation distribution, radial in this case, acquires a different distribution after traversing a chiral medium, which is directly determined from the Stokes parameters. The proof-of-concept was demonstrated by simultaneously measuring the concentration of two unmissable enantiomeric solutions, L-proline and (R)-(+)-Limonene. This novel technique paves the way towards accurate monitoring of more complex samples, for example, the monitoring of chiral compound synthesis or the quantification of organic compound concentrations in the atmosphere.

\subsection{Applications in Optical Tweezers}
Optical tweezers, conceived more than 50 years ago by the 2018 Physics Nobel Prize \red{winner} Arthur Ashkin \cite{Ashkin1970}, provide a powerful and versatile tool for micro- and nano-structures manipulation, with potential impact in various applications. The force required to perform optical manipulation arises from either the linear or angular momentum of light \cite{Padgett2011}. The latter can be in the form of spin (associated with light's polarisation) \cite{Poynting1984,Beth1936} or orbital (associated with light's wavefront) \cite{Allen1992} angular momentum. In particular, SLMs enable the simultaneous manipulation of multiple particles in 2D and 3D configurations or along exotic trajectories. See for example \cite{yang2021,Otte2020tweezers} for nice reviews about using structured light in optical tweezers. Even though only scalar beams with homogeneous polarisation were initially considered, vector beams provide additional capabilities to further advance the existing manipulation techniques. For example, \red{in 2009} it was demonstrated that vector beams with radial polarisation can increase the axial trapping efficiency compared to scalar beams of linear polarisation\cite{Michihata2009}. Other studies have shown that vector beams with azimuthal polarisation possess stronger lateral trapping forces compared to those with radial polarisation \cite{Bhebhe2018}. A typical implementation of optical tweezers with vector beams is schematically represented in \red{Fig.~\ref{fig:applications} (e)}. Structured light beams have also enabled the generation of tractor beams, in which the optical force exerted on the micro- or nano-particles produces an acceleration in a direction contrary to the photon flow capable of dragging small microparticles towards the photons source. The combination of cylindrical vector beams with the photophoretic effect has also enabled a long-distance, stable, and switchable optical transport, which can pull or push the particles along the optical beam axis, by alternating between vector beams with radial or azimuthal polarisation\red{, as demonstrated in 2013}\cite{Brzobohat2013}. As such, by modifying the intensity, wavefront, and polarisation distributions of light beams it is now possible to transport particles along open and closed trajectories, make them rotate, and trap absorbing and non-absorbing particles. In other words, the ability to tailor the various properties of light has enhanced the capabilities of optical tweezers, thus establishing the bases for their next generation.

\section{Looking ahead}

The field of structured light is moving fast in multiple directions. In its very short existence time, it has gained popularity across various fields, where its fascinating properties have \red{inspired novel concepts and} provided an alternative tool for developing innovative applications and improving current ones. \red{In the field of optical manipulation, structured light has enabled not only manipulating several particles using multiple vector beams with independent properties\cite{Bhebhe2018}, but also developing a vectorial tractor beam capable of pulling and pushing microparticles in free space by simply switching from an azimuthal to a radial vector beams \cite{Brzobohat2013}. These two examples exemplify their full potential has not yet been exploited, thus indicating the following years will reveal more fascinating optical manipulation techniques. By way of example,} it is expected that other schemes will be unveiled, namely long-range control of microparticles with optical needles, the complete control of chiral molecules using, for instance, spatially disjoint vector beams \cite{Medina-Segura:2023,HuXiaobo2021} \red{or self-focusing vector beams \cite{Hu2023}}, and manipulating particles in the presence of turbulence with resilient structured light beams \cite{nape2022}.

\red{In optical communications, several pioneering works have proposed potential solutions based on structured light to overcome the pending bandwidth capacity crunch. For example, solutions include using scalar beams and vector beams in both free-space and optical fibre. Many of these techniques have also been extended to underwater optical communications \cite{Zhao2017OL}. For instance,} it has allowed a new coding scheme for optical communications, where the information channels are defined in terms of the concurrence \cite{Singh2023}. \red{This approach takes advantage of the invariance of vector beams upon propagation through complex media, one of the properties of vector beams yet to be fully exploited. Another possible direction is utilising partially coherent structured beams since pioneering works have demonstrated that they are highly resistant to the deleterious effects of atmospheric turbulence in free-space propagation \cite{Gbur2014,Wang2017}. This constitutes one of the reasons why partially coherent beams started to be explored in various coordinate systems \cite{Yepiz2020,Perez-Garcia_2022}.}

\red{At the quantum level, structured light also provides the means towards more secure quantum communication links. Of particular relevance are quantum key distribution (QKD) protocols, which have benefited from the implementation with higher dimensions, thus increasing their security in proportion to the dimension of the encryption basis used \cite{Otte2020, Mafu2013, Mirhosseini2015}. While these have been implemented with beams endowed with OAM, it is possible that in the following years, other types of beams, such as Ince- or Mathieu-Gaussian beams exploring properties like their ellipticity, can be utilised for developing novel security protocols. In addition,} the realisation that entanglement decay is identical in classical and quantum systems, opened the possibility of characterising quantum links in real-time \cite{Ndagano2017}. \red{In particular, the idea of implementing a hybrid classical-quantum optical channel has been suggested as a possible direction \cite{forbes2022method}.}

In optical metrology, while diverse techniques have also been proposed, the search for improved methods that require fewer measurements (ideally one), show higher precision, are robust against environmental factors, or are easier to implement, sets possible research directions. \red{Here, it is very likely that vector beams featuring non-homogeneous polarisation distribution will pioneer the next generation of polarisation-based applications, as has been already suggested\cite{Toppel2014}. For instance, to measure the physical properties of solids or liquids in two dimensions, as was recently demonstrated \cite{Hu2024}.} 

Furthermore, the demonstration of novel types of structured light beams, as well as the development of novel generation and characterization techniques advance hand by hand with the applications development. Sometimes inspired by a specific application need, and others inspiring novel ones. Perhaps one of the biggest challenges is related to the real-time characterisation of vector beams, especially for metrology applications where the system under study changes dynamically with time. Along this direction, a few steps have already been given towards a real-time Stokes polarimetry (see for example \cite{Zhao2019,Cox2023}). Another example where characterisation methods are needed is structured light beams with intensity or polarization distributions that change upon propagation. For instance, rotating petal beams have been used to measure refractive indexes\cite{Dorrah2018}. In addition, the change in the “vectorness'' of spatially disjoint vector beams, such as the Parabolic-Gauss vector beams \cite{HuXiaobo2021}, which the VQF is not sensitive to, is an open problem with few solutions addressing it, being the Hellinger distance a proposed approach\cite{Aiello2022}. Therefore, additional characterisation techniques to meet these challenges are key to expanding the influence of structured light, particularly considering that they could be useful for designing alternative metrics for specific applications.

\red{Since} computer-controlled devices offer great versatility and performance for generating scalar and vector beams, they are usually best suited for proof-of-principle demonstrations due to their limitations in terms of, mainly, bit-depth and cost. Alternative approaches are then needed for the transition from laboratory demonstrations to real-world devices, which constitute the ultimate goal in applied research. This requirement can be approached from different angles. For instance, ongoing research focuses on generating structured light beams using compact or more affordable devices, such as planar refractive elements or printed holograms \cite{Torres-Leal2023,PlannarOptics}, a research direction that has already given some fruits \cite{YuanChen2020,PanFu2023,Cisowski2023}. The fabrication of improved metasurfaces for the generation of beams with higher quality is also gaining track, see for example the recently proposed metasurface to implement complex amplitude modulation \cite{DeOliveira2023}. Alternative solutions propose the generation of structured light beams directly from laser cavities, showing remarkable potential as it eliminates problems typically associated with their indirect generation, such as lower conversion efficiencies \cite{Naidoo2016}. 

\red{Finally, the introduction of Artificial Intelligence (AI) will be key in advancing all areas of the field. For instance, neural-networks-based methods are taking over conventional techniques in detecting structured light beams, a research direction that is evolving very rapidly \cite{Bai2023, Hu2023NeuralNetwork, cox2022interferometric, ragheb2020identifying, zhang2024spatial}. Of particular relevance is the possibility of identifying beams propagating in non-ideal conditions, such as atmospheric turbulence\cite{du2023recognition, avramov2022classifying}. AI is also being used for application development, such as improving high-dimensional QKD protocols with deep learning algorithms, a research line that has already started \cite{wang2021integrating}.}

As a closing remark, we would like to highlight that while the past two decades have seen tremendous advancements in the structured light field, there is more to come, especially on the applications side. We anticipate many more applications in addition to the ones we considered here will be proposed, permeating other research fields and enabling applications that are challenging with conventional techniques but possible with this new tool. Hence, it is very likely that we will witness structured light becoming a mature field providing novel solutions to new and old problems.\\

\noindent
{\bf Conflict of Interest} The authors have no conflicts to disclose.\\

\noindent
{\bf Data availability} Data sharing does not apply to this article as no new data were created or analysed in this study.
\section*{References}
\input{References.bbl}
\end{document}

%% file: References.bbl
%